\documentclass[9pt,preprint2,natbib209]{aastex61}
\usepackage{graphicx}% Include figure files
\usepackage{dcolumn}% Align table columns on decimal point
\usepackage{bm}% bold math

\def\apjs{Astrophys. J.  Supp. Seri.}

\def\apjl{Astrophys. J. Lett.}

\def\aap{Astro. \& Astrop.}

\begin{document}


\title{Effects of Irradiation on the Evolution of Ultra-compact X-ray Binaries.}

\correspondingauthor{Guoliang L\"{u}}\email{guolianglv@xao.ac.cn}

\author{Guoliang L\"{u}}
\affil{School of Physical Science and Technology,
Xinjiang University, Urumqi, 830046, China}

\author{Chunhua Zhu}
\affil{School of Physical Science and Technology,
Xinjiang University, Urumqi, 830046, China}

\author{Zhaojun Wang}
\affil{School of Physical Science and Technology,
Xinjiang University, Urumqi, 830046, China}

\author{Hoernisa  Iminniyaz}
\affil{School of Physical Science and Technology,
Xinjiang University, Urumqi, 830046, China}

%\altaffiltext{3}{Zentrum f\"{u}r Astronomie, Institut f\"{u}r
%Theoretische Astrophysik, Universit\"{a}t Heidelberg,
%Albert-\"{U}berle-Str. 2, D-69120 Heidelberg, Germany.}

\date{\today}

%\pagerange{\pageref{firstpage}--\pageref{lastpage}} \pubyear{2007}

%\maketitle

%\label{firstpage}

\begin{abstract}
Using the Modules for Experiments in Stellar Astrophysics code,
we investigate the influences of irradiation on ultra-compact X-ray binary (UCXB) evolution.
Although the persistent UCXBs have short orbital periods which result in
high irradiation flux, the irradiation hardly affects the evolution of
persistent sources because the WDs in these binaries have large masses which lead to very
low irradiation depth.
The irradiation has a significant effect on the transient sources during outburst phase.
At the beginning of the outburst, high X-ray luminosity produces high radiation flux, which results in the
significant expansion of WD. Then, the irradiation triggers high mass-transfer rates, which can last
several days for the transient sources with WDs whose masses are larger than $\sim0.015 M_\odot$ or
several hundred years for these sources with WDs whose masses are less than $\sim0.012 M_\odot$.
The observed three persistent UCXBs, XTE J0929-314, 4U 1916-05 and SWIFT J1756.9-2508,
may belong to the latter.

\end{abstract}

%\maketitle
\section{Introduction}

Ultra-compact X-ray binaries(UCXBs) are a subclass of low-mass X-ray binaries (LMXBs).
They are characterized by an orbital period of less than 60 minutes in which a neutron star (NS) or black
hole is accreting matter from its companion star.
This indicates that the companion star must fill its Roche lobe.
Therefore, due to the short orbital period, the donors star in UCXBs must be hydrogen deficient  such as white dwarfs (WDs) or
helium stars\citep{Paczynski1981,Nelson1986}.
Up to now, there are about 30 known UCXBs or candidates,
and 13 of them have been observed orbital periods\citep{Zand2007,Liu2007,Nelemans2010}.
UCXBs may be produced by tidal captures or direct collisions in the globular clusters\citep[e.g.,][]{Verbunt1987,Ivanova2010}, or may
originate from binary systems which undergo complex interactions such as mass exchange, common envelope,
gravitational radiation\citep[e.g.,][]{Yungelson2002,Belczynski2004,Zhu2012}. Recently, \cite{Chen2016} suggested
that UCXBs may originate from intermediate-mass X-ray binaries driven by magnetic braking of stars with
strong magnetic field (100-10000G).

\cite{Yungelson2008} and \cite{Haaften2012a} described
the details for the evolution of UCXBs with helium star and WDs, respectively. They obtained similar results: With the evolution of UCXBs,
the donors' masses and the mass-accretion rates of compact objects become lower and lower, while their orbital periods
widen. Finally, UCXBs become binaries with low mass ratios and orbital periods of $\sim$70-80 minutes\citep{Haaften2012a}.
Having used data from the RXTE All-Sky Monitor, \cite{Haaften2012b} studied the long-term X-ray luminosity behavior of 14
UCXBs. They found that the UCXBs with orbital periods longer than about 50 minutes are much brighter than
theoretically estimated values in \cite{Haaften2012a}. Very recently, \cite{Sengar2017} obtained similar results.
A possible explanation is that
these pulsars irradiate their low-mass companion stars \citep{Haaften2012b}. The influences of irradiation in
semi-detached compact binaries have been discussed
by many literatures\citep[e. g.,][]{Podsiadlowski1991,Hameury1993,Buning2004}. The one of important influences is
that the irradiation drives higher mass transfer. However, these literatures focus on main sequence stars or
sub-giant stars as irradiated stars. To our knowledge, irradiated WDs in UCXBs are seldom referred to.

In this work, we investigate the effects of irradiation on WDs in UCXBs, and discuss the evolution
of UCXBs.

\section{Model}
UCXBs are composed of a NS and a WD. In our
work, NS is a mass point with $1.4 M_\odot$, and
we do not simulate its evolution. We focus on
the effects of the irradiation on WDs and the all consequences for
UCXBs. In this work, we
use the Modules for Experiments in Stellar Astrophysics code
(MESA, see \cite{Paxton2011,Paxton2013,Paxton2015} for details.) to
simulate the evolution of irradiated WDs and UCXBs.

\subsection{WD Model}
\begin{figure}
\includegraphics[totalheight=3.0in,width=3.0in,angle=-90]{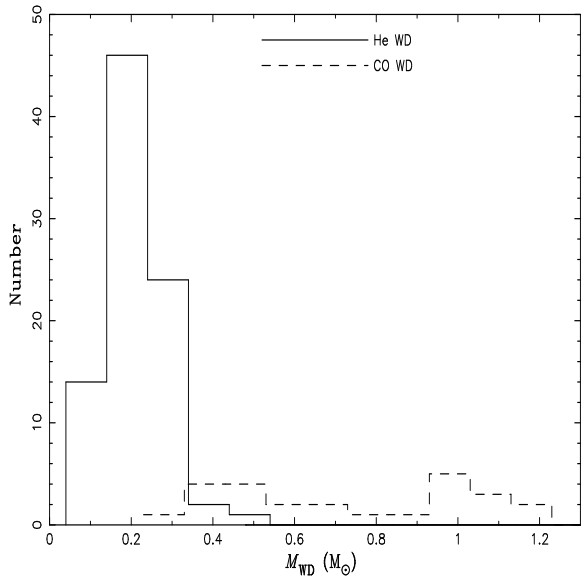}
\caption{Distributions of WDs' masses for
 120 pulsars with WDs as their companion stars. He and CO mean
 that WD is rich helium or carbon-oxygen, respectively. Median companion mass, assuming $i=60$ degrees,
 is taken as WD mass.
 Data come from \cite{Manchester2005}. }
\label{fig:mass}
\end{figure}

The donors in UCXBs may be helium stars or WDs. \cite{Yungelson2008} investigated
the evolution of low-mass helium stars in semi-detached binaries, and showed that
their evolution is similar with these of WDs after these helium stars
undergo a phase of high mass transfer\citep[Also see][]{Haaften2012b}.
The compositions of these WDs maybe rich helium or rich carbon-oxygen.
\cite{Haaften2012b} gave the compositions of the donors in 10 UCXBs.
About a half of these donors are rich helium.
In all known radio pulsars, there are 120 pulsars with WDs as their companion stars\citep{Manchester2005}.
Figure \ref{fig:mass} gives the distributions of WDs' masses for these pulsars.
Among them, about 77\% are He WDs, and their masses are $\sim$ 0.2 $M_\odot$.
Although the majority of these binaries can not evolve into
UCXBs in Hubble time, WDs' masses and compositions may be similar with these in progenitors of UCXBs.
Therefore, for simplicity, in this work, we only focus on He WDs as donors in UCXBs. Hereafter,
WD means He WD unless a special note.

\begin{figure}
\includegraphics[totalheight=3.0in,width=3.0in,angle=-90]{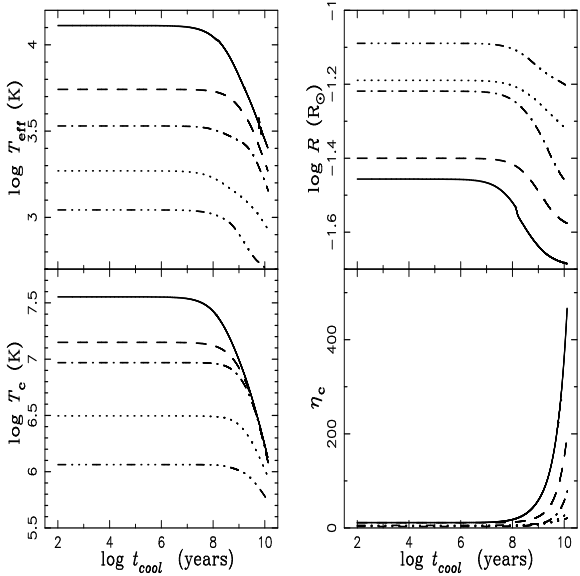}
\caption{The evolution of the effective temperature ($T_{\rm eff}$),
radius ($R$), central temperature ($T_{\rm c}$) and central degeneracy ($\eta_{\rm c}$),
calculated by MESA, for single WDs with different masses in Hubble time. The solid, dashed, dash-dotted, dotted and
dash-dot-dot-dotted lines represent the WDs with masses of 0.2, 0.1, 0.05, 0.02 and 0.01 $M_\odot$,
respectively.}
\label{fig:wdlum}
\end{figure}

\begin{figure}
\includegraphics[totalheight=3.0in,width=3.0in,angle=-90]{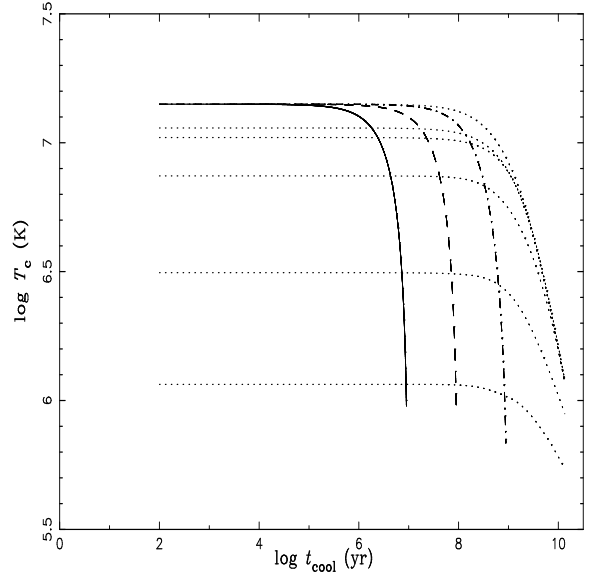}
\caption{The evolution of the central temperature ($T_{\rm c}$) of a
0.1 $M_\odot$ WD with different mass-loss rates.
Solid, dashed and dash-dotted lines represent the mass-loss rates of
$10^{-8}, 10^{-9}$ and $10^{-10} M_\odot$ yr$^{-1}$, respectively. The
dotted lines are the evolution of $T_{\rm c}$ for different mass WDs. From the
top to the bottom, WDs' masses are 0.1, 0.08, 0.06, 0.04, 0.02 and 0.01 $M_\odot$, respectively.
}
\label{fig:mltc}
\end{figure}

Although WDs are the remnants of low-mass stars, they also have the evolutionary tracks.
Figure \ref{fig:wdlum} shows the evolution of the effective temperature ($T_{\rm eff}$),
radius ($R$), central temperature ($T_{\rm c}$) and central degeneracy ($\eta_{\rm c}$)
for WDs with different masses. For the evolution of UCXBs, WD's radius is
critical factor. \cite{Deloye2003} investigated the mass-radius relation of WD donors in UCXBs, and found that
this relation depends on the central temperatures of WDs.
In Figure \ref{fig:wdlum}, $T_{\rm c}$ of WD whose mass is higher than 0.02 $M_\odot$ hardly
cools down lower than $10^6$ K within 13 Gyr. However, the WDs in UCXBs are undergoing the mass loss, and
have different evolutionary tracks. Figure \ref{fig:mltc} gives the evolution of $T_{\rm c}$ for a
0.1 $M_\odot$ WD with different mass-loss rates.
Obviously, $T_{\rm c}$ rapidly decreases with WD mass reducing in a timescale very shorter than Hubble time.
The main reason is as follows: Mass loss results in the decreases of gravitational potential. The thermal energy
leads to the expansion of WD, which gives rises to the fall of $T_{\rm c}$.

\begin{figure}
\includegraphics[totalheight=3.0in,width=3.0in,angle=-90]{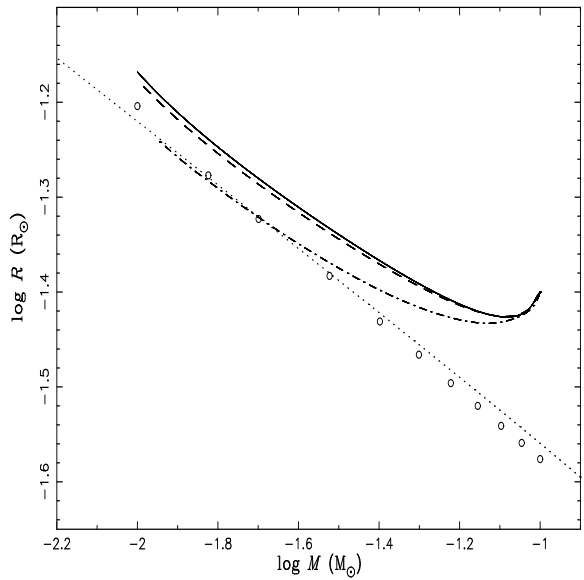}
\caption{Mass-radius relations of WDs. The circles represent the radii of
WDs after 13 Gyr cooling, which is calculated by $MESA$ code.
 Solid, dashed and dash-dotted lines represent
the radii's evolution of 0.1 $M_\odot$ WD with mass-loss rates of $10^{-8}, 10^{-9}$
and $10^{-10} M_\odot$ yr$^{-1}$, respectively.
Dotted line shows the mass-radius relation of zero-temperature WDs by Eq. (\ref{eq:mlr}). }
\label{fig:mlr}
\end{figure}

Figure \ref{fig:mlr} gives the mass-radius relations of 0.1 $M_\odot$  WD with the different mass-loss rates.
Obviously, with WD mass reducing, its radius closes to the radius of zero-temperature WD.
Therefore, $T_{\rm c}$ of WD only determines the beginning of UCXB, and it mainly depends on
the mass-loss rates of WD in UCXB.
Here, the mass-radius relation of zero-temperature WDs can be approximated by
\begin{equation}
R_{\rm WD}=0.0115\sqrt{(M_{\rm CH}/M_{\rm WD})^{2/3}-(M_{\rm WD}/M_{\rm CH})^{2/3}},
\label{eq:mlr}
\end{equation}
where $M_{\rm WD}$ is the mass of WD, and $M_{\rm CH}=1.44 M_\odot$ is the Chandrasekhar mass \citep{Tout1997}.
Eq.(\ref{eq:mlr}) is an approximate fitting formula.
For a sufficiently cool white dwarf (k$T_{\rm c} <<$ degenerate energy),
\citep{Tout1997} used Peter Eggletton¡¯s stellar evolution code to calculate the radii of WDs,
and finally gave Eq.(\ref{eq:mlr}) by fitting these WD¡¯s radii and masses.
Because Eq.(\ref{eq:mlr}) does not depend on temperature, we call it as the mass-radius relation of zero-temperature WDs.
In fact, the radius of WD depends on mass, temperature and composition.
In our work, the radii of WD may be lower than the zero-temperature line for certain WD masses.

\subsection{Evolution of Orbital Angular Momentum and Mass Transfer}
The change of orbital angular momentum is the key to understand the evolution of UCXBs.
In binary systems, mass variations (including mass loss, mass transfer), tide,
gravitational radiation and magnetic braking can change orbital angular momentum.
\cite{Hurley2002} gave careful program for it, which is adopted by \cite{Mink2013}.
In MESA, the above all physical processes are considered\citep{Paxton2015}.

For the binary systems composed of a NS and a WD, the gravitational radiation
determines the evolution of orbital angular momentum before the WD fills its Roche lobe.
After that, the mass transfer dominates the orbital change.
However, the mass-transfer rate ($\dot{M}_{\rm RL}$) from the WD donor and
the mass-accreting rate ($\dot{M}_{\rm }$) of the NS accretor are unclear.

For the former, considering the finite scale height ($H_{\rm P}$) of the donor's atmosphere, \cite{Ritter1988}
suggested mass-transfer rate by
\begin{equation}
\dot{M}_{\rm RL}=\dot{M}_{\rm 0}\exp\left({\frac{\Delta R_{\rm L}}{H_{\rm P}}}\right),
\label{eq:mesa}
\end{equation}
where $\dot{M}_0$ is a parameter depending on donor's structure, and $\Delta R_{\rm L}=R_{\rm donor}-R_{\rm L}$.
Here,  $R_{\rm L}$ is the Roche lobe radius of the donor.
Following \cite{Ritter1988} and
taking the detailed structure of the donor's outer layers, \cite{Kolb1990} gave a formula for
$\dot{M}_{\rm RL}$. The details can be seen in \cite{Paxton2015}.

According to the hydrostatic and thermal equilibria of a star,
\cite{Paczynski1972} gave a dynamical estimate by
\begin{equation}
\dot{M}_{\rm RL}=\frac{M_{\rm donor}}{P_{\rm orb}}\left(\frac{{\Delta R_{\rm L}}}{R_{\rm L}}\right)^{n+3/2},
\end{equation}
where $P_{\rm orb}$ is the orbital period and $n$ is the polytropic
index of the donor. Similarly, using the stellar evolutionary code originally developed by
\cite{Eggleton1971,Eggleton1972,Eggleton1973} (This
code is noted by PPE code in this work),
\cite{Han2002} calculated it by
\begin{equation}
\dot{M}_{\rm RL}={\rm C}\max\{0,\left(\frac{{\Delta R_{\rm L}}}{R_{\rm L}}\right)^{3}\},
\label{eq:ppe}
\end{equation}
where $C$ is a constant, and it is taken as $1000 M_\odot$ yr$^{-1}$.

In a rapid binary star evolutionary (BSE) code,
\cite{Hurley2002} calculated $\dot{M}_{\rm RL}$ by
\begin{equation}
\dot{M}_{\rm RL}=3\times10^{-6}[\min(5,M_{\rm donor})]^2[{\rm ln}(R_{\rm donor}/R_{\rm L})]^3M_\odot {\rm yr^{-1}}.
\label{eq:bse}
\end{equation}
\cite{Mink2013} used the above formula to calculate the mass-transfer rates in the binary systems.

The above three mass-transfer rates depend on $\Delta R_{\rm L}$ via exponent, power and logarithm functions, respectively.
Based on mathematic knowledge, Eq. (\ref{eq:mesa}) gives the highest sensibility for
the mass-transfer rate depending on $\Delta R_{\rm L}$, while Eq. (\ref{eq:bse})
gives the lowest one.
In order to discuss the effects of the different mass-transfer rates on the evolution of UCXBs,
we use Eqs. (\ref{eq:mesa}) and (\ref{eq:bse}) to calculate $\dot{M}_{\rm RL}$, respectively.

Similarly, the accretion efficiency, $\beta=\frac{\dot{M}_{\rm }}{\dot{M}_{\rm RL}}$,
is also hardly determined due to the strong magnetic fields of NSs
and the radiation pressure of X-ray luminosity. Furthermore, the orbital angular momentum
carried by the lost matter also is unclear, which mainly is relative to the vicinity of the
mass lost\citep{Tauris2006}.
Following \cite{Podsiadlowski2002}, we take $\beta=0.5$ and
assume that the mass is lost from the vicinity of the accreting NS.

\subsection{Irradiation Model}
In the close binaries, irradiation can affect the evolution of donors, even binary systems.
\cite{Podsiadlowski1991} investigated that the irradiation drives the mass transfer in LMXBs.
\cite{King1995} and \cite{Buning2004} developed the irradiation feedback
model for compact binaries, and they found that the mass transfer rates become
unstable and these binaries experience mass transfer cycles.
MESA uses the irradiation model provided by \cite{Guillot2010}, in which an analytical approach
is used to simulate stellar atmospheres.
In MESA, the irradiation flux ($f^0_{\rm irr}$) and the irradiation column depth ($r_{\rm irr}$)
are important input parameters.
The former can be given by
\begin{equation}
f^{0}_{\rm irr}=\alpha_{\rm irr}\frac{L_{\rm X}}{4\pi a^2},
\end{equation}
where $\alpha_{\rm irr}$ is the irradiation efficiency, and $L_{\rm X}$ is the X-ray luminosity emitted by accreting NS and
$a$ is the binary separation. In LMXBs, $\alpha_{\rm irr}\sim0.01 - 0.1$\citep{Stevens1992,Buning2004,Benvenuto2012}.
In this work, $\alpha_{\rm irr}=0.01$.

The $L_{\rm X}$ is determined by not only the mass-accretion rate but also thermal disk instability. The later divides the UCXBs
into persistent or transient sources, which depends on the mass-accretion rate. UCXB is a transient X-ray source if
the mass-accretion rate is lower than a certain critical value, $\dot{M}_{\rm c}$, or else it is a persistent source.
Following \cite{Zhu2015}, we use the formula in \cite{Dubus1999} and \cite{Menou2002}
to calculate $\dot{M}_{\rm c}$ for the hydrogen-rich and heavier element disks, respectively. For the persistent X-ray sources,
$L_{\rm X}$ can
be approximated by
\begin{equation}
L_{\rm X}=\eta \dot{M}{\rm c}^2\simeq5.7\times10^{35}(\frac{\eta}{0.1})(\frac{\dot{M}}{10^{-10}M_\odot{\rm yr^{-1}}}) {\rm erg\ s^{-1}}
\end{equation}
where $\eta=0.1$ is the efficiency of converting accreted mass into
X-ray photons. For the transient X-ray sources, following \cite{Belczynski2008}, we assume that $L_{\rm X}=0$ during the quiet phase and
$L_{\rm X}=0.1 L_{\rm Edd}$ during the outburst phase, where $L_{\rm Edd}$ is the Eddington luminosity.

\begin{figure}
\includegraphics[totalheight=3.0in,width=3.0in,angle=-90]{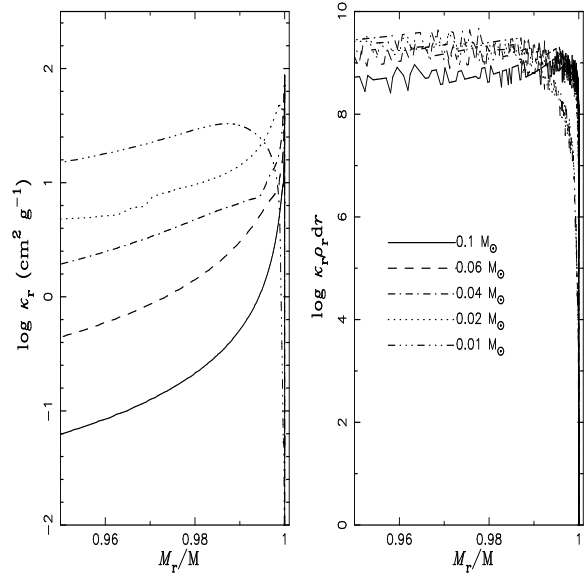}
\caption{The logarithm of the opacity ($\kappa$) and $\kappa \rho {\rm d}r$ around the stellar surface for
the low-mass WDs. The different line styles represent WDS with the different masses which are
noted in the right panel. }
\label{fig:opacity}
\end{figure}

The later ($r_{\rm irr}$) is the penetrating depth of $f_{\rm irr}$ below the photosphere. It depends on the irradiation
spectra and the local physical and chemical conditions of the WD donor. For simplicity, we assume that
$f_{\rm irr}$ penetrating in the star decreases exponentially as $f^{0}_{\rm irr}{\rm e}^{-\int\kappa \rho {\rm d}r }$£¬
where $\kappa$ and $\rho$ are the local opacity and mass density of WD, respectively, and $r$ is the distance from the
WD center.
Simultaneously, we also assume that $r_{\rm irr}=r$ where $f_{\rm irr}=f_{\rm int}$, where $f_{\rm int}$ is the
intrinsic flux and equals $\frac{L_{\rm r}}{(4\pi\ r^2)}$. Here, $L_{\rm r}$ is the intrinsic luminosity of WD at $r$.
Thus it can be seen that $r_{\rm irr}$ is mainly determined by ${-\int\kappa \rho {\rm d}r }$.
Figure \ref{fig:opacity} gives the values of log $\kappa$ and log $\kappa \rho {\rm d}r$ around the stellar surface for
the WDs with different masses. Obviously, $f_{\rm irr}$ can only penetrate into very thin layer around WD surface.
The larger WD mass is, the more difficult the penetration is. Therefore, the irradiation only results in the
heating and radial expansion of low-mass WD.

%Simultaneously, we also assume that $r_{\rm irr}=r$ where $f_{\rm irr}=\frac{f^{0}_{\rm irr}}{10000}$, that is,
% $\int^{r_{\rm irr}}_{R_{\rm WD}}\kappa \rho {\rm d}r={\rm ln\ 10000}$.

\section{Results}%Influences of Irradiation on Evolution of UCXBs

\begin{figure}
\includegraphics[totalheight=3.0in,width=3.0in,angle=-90]{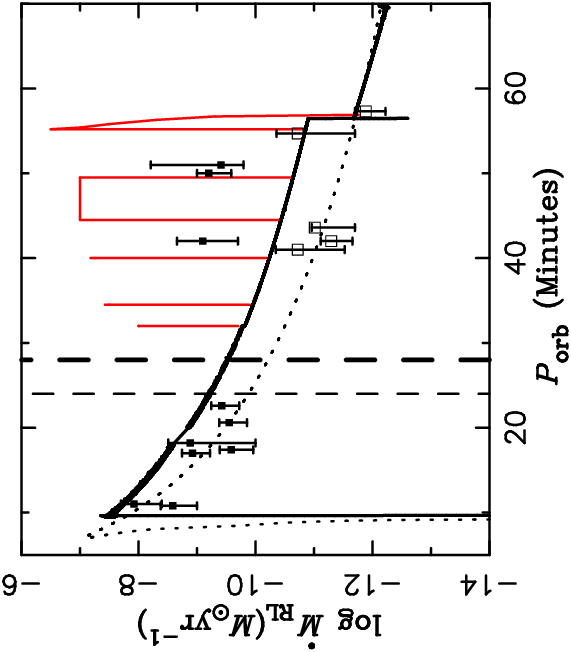}
\caption{The evolution of UCXBs with different mass-transfer rates.
The solid and dash lines represent the different mass-transfer rates
described by Eq. (\ref{eq:mesa}) and (\ref{eq:bse}), respectively.
The solid and empty squares give the observational data for
persistent and transient UCXBs, respectively. The data come from \cite{Heinke2013} and \cite{Cartwright2013}.
The thick and thin dashed lines represent the orbital period thresholds for the thermalviscous instability
for the mass-transfer rates of Eqs. (\ref{eq:mesa}) and (\ref{eq:bse}), respectively.
The red solid lines show the evolution of the WDs irradiated in
transient UCXBs during outbursts. From the left to the right, the irradiated WDs are
0.02, 0.018, 0.015, 0.012 and 0.01 $M_\odot$, respectively.}
\label{fig:pm}
\end{figure}
Based on the assumptions in the above section, we
simulate the evolution of a binary system composed of
a NS with $1.4 M_\odot$ and a WD with $0.1 M_\odot$.
Because most of the progenitors of UCXBs undergo common envelope evolution,
it is reasonable to assume a circular orbit.
The orbital period is
3 hours when the WD is born in this systems. After the cooling of 2 Gyr,
the WD fills its Roche lobe due to the gravitational wave radiation.
The binary system evolves into a UCXB.

\subsection{Evolution of UCXBs}
Figure \ref{fig:pm} shows the evolution of this binary system as UCXB with different mass-transfer rates.
The mass transfer in the model with the mass-transfer rate calculated by Eq. (\ref{eq:mesa}) from \cite{Ritter1988}
can occur even if the WD is just closed to its Roche lobe. Therefore, compared with the model with the mass-transfer
rate calculated by Eq. (\ref{eq:bse})\citep{Hurley2002}, it has longer orbital period at the beginning of UCXB phase.
For the same orbital period, the mass-transfer rate in the former is higher than the one in the latter.
When matter transfers from WD to NS, the orbital period becomes wider and wider.
The mass-transfer rate calculated by Eq. (\ref{eq:mesa}) rapidly reduces at $P\sim 57$ minutes because
the widen of orbital period produced by high mass-transfer rate exceeds the expansion of WD due to its mass decreases.
Based on the X-ray luminosities of UCXBs in \cite{Cartwright2013}, \cite{Heinke2013}
estimated their mass-transfer rates, which are showed by the squares in Figure \ref{fig:pm}.
To be exact, these estimated values should be the mass-accretion rates of NSs.
Although the mass-transfer rate of Eq. (\ref{eq:bse}) agrees with the observational values
and Eq. (\ref{eq:mesa}) give higher mass-transfer rate, we can not conclude that
which is better because the efficiency of mass-accretion, $\beta$, is unclear.

\cite{Haaften2012b} calculated the orbital period threshold for the thermalviscous instability of helium accretion disks in UCXBs,
and found that the threshold is about 28 minutes.
\cite{Sengar2017} also computed this threshold. It is about 22 minutes for the helium accretion disk under X-ray heating.
 According to \cite{Menou2002} and considering that a mass-transfer rate corresponds to a orbital period in UCXBs,
we calculate the orbital period thresholds  with
different mass-transfer rates(Eqs. (\ref{eq:mesa}) and (\ref{eq:bse})),
which are represented by thick and thin dash lines in Figure \ref{fig:pm}, respectively.
Based on the formula in \cite{Menou2002},
the longer the orbital period is, the higher the critical mass transfer rate is.
Therefore, the orbital period threshold for the thermal disc instability
from Eq. (\ref{eq:mesa}) is longer than that from Eq. (\ref{eq:bse}).
Correspondingly, the critical mass transfer rate in the former also is higher than that in the latter.
Our results are consistent with previous works.

On the observations, UCXBs whose orbital periods are shorter than 30 minutes are persistent,
which is consistent with theoretical estimates. However, there are some persistent sources (XTE J0929-314, 4U 1916-05 and SWIFT J1756.9-2508) for
UCXBs with orbital periods between 40 and 60 minutes. Theoretically, they should be transient.
\cite{Haaften2012b} suggested that the WDs in these UCXBs are heated by the irradiation from
X-ray emitted by accreting NS. \cite{Sengar2017} considered that these systems may be LMXBs which are evolving into
UCXBs. However, these LMXBs have mass-transfer rates lower than these of UCXBs (See Figure 2 in \cite{Sengar2017}).

\subsection{Influences of Irradiation on UCXBs' Evolution }
The most remarkable feature  of UCXBs is short orbital periods. The X-ray  produced by
accreting NS strongly irradiates WD.
This work focuses on the influences of irradiation on UCXBs' evolution.
The conditions of irradiation in the persistent and the transient UCXBs are different.
In our work, the orbital period thresholds for the thermalviscous instability of helium accretion disks
are about 22 and 28 minutes for the mass-transfer rates of Eqs. (\ref{eq:bse}) and (\ref{eq:mesa}), respectively.
They are given by thin and thick dashed lines in Figure \ref{fig:pm}.
Correspondingly, the masses of WDs in the persistent UCXBs is larger than $\sim 0.025 $ and $\sim 0.018 M_\odot$, respectively.
For simplicity, in the present paper, we assume that WDs in the persistent UCXBs have mass larger than $0.02 M_\odot$.

\begin{figure*}
\begin{tabular}{lr}
\includegraphics[totalheight=3.0in,width=3.2in,angle=-90]{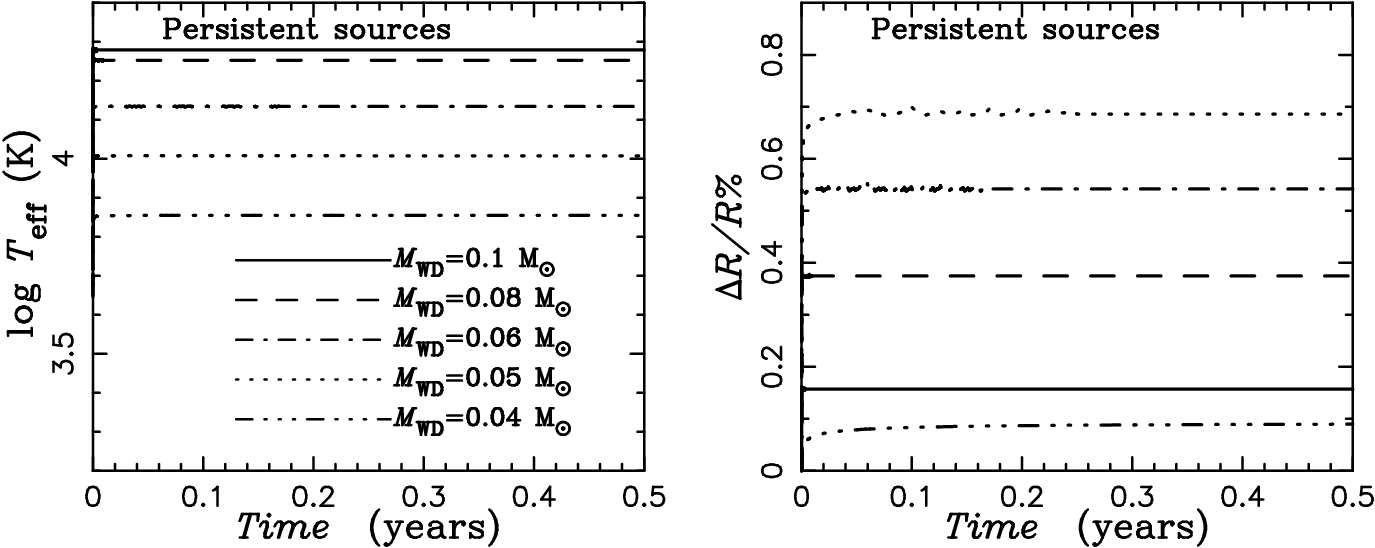}&
\includegraphics[totalheight=3.0in,width=3.2in,angle=-90]{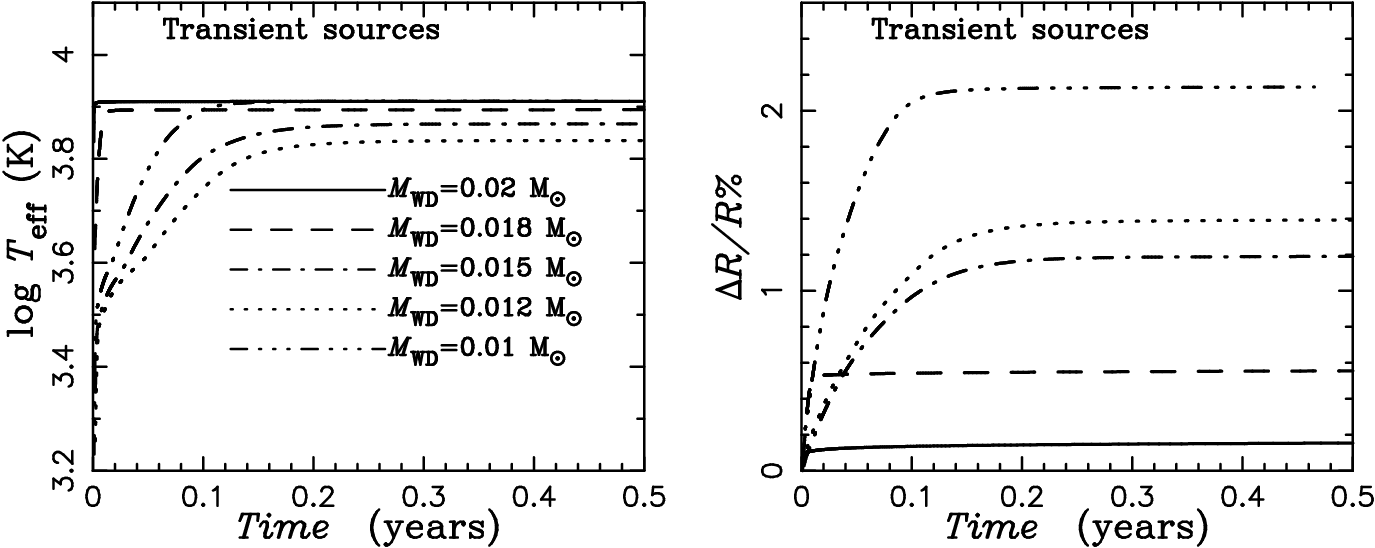}\\
\end{tabular}
\caption{The varieties of the effective temperatures ($T_{\rm eff}$)
and the relative radii ($\Delta R/R$) for irradiated WDs.
The 'Time' of the  x-coordinate means the time since the start of irradiation.
The different line styles represent WDs with the different masses which are
noted in the middle zone of the left panel.
}
\label{fig:itev}
\end{figure*}

Table \ref{tab:irr} gives the physical parameters for irradiation model. Figure \ref{fig:itev} shows
the varieties of the effective temperature ($T_{\rm eff}$)
and the relative radius ($\Delta R/R$) after WDs are irradiated.
For the persistent UCXBs ($M_{\rm WD}>0.02 M_\odot$), WDs rapidly reach a new thermodynamic equilibrium
within less than 1 s ($M_{\rm WD}=0.1 M_\odot$) and about 100 days ($M_{\rm WD}=0.04 M_\odot$) due to
high irradiation flux ($f_{\rm irr}^0$) and small irradiation depth ($r_{\rm irr}$). However,
the increase of WD radius is too small to affect UCXB evolution. Even though we increase
$f_{\rm irr}^0$ via varying $\alpha_{irr}$ from 0.01 to 0.1, the increase of WD radius is very small
due to small $r_{\rm irr}$.

For the transient UCXBs ($M_{\rm WD}<0.02 M_\odot$), X-ray luminosity of the accreting NS does not
depend on the mass-transfer rate. During the quiet phase, X-ray luminosity ($\sim 10^{31}-10^{32}$ erg s$^{-1}$)
is very low, and we do not consider irradiation effect. During the outburst phase,
X-ray luminosity can rise to about 0.1$L_{\rm EDD}$, which is taken as $10^{37}$ erg s$^{-1}$ in
this work (See Table \ref{tab:irr}). As showed in Table \ref{tab:irr} and Figure \ref{fig:itev},
compared with the WDs in the persistent UCXBs, WDs in the transient UCXBs have larger irradiation depth.
Therefore, the timescale of WD reaching a new thermodynamic equilibrium becomes long and the variety of relative radius
increases.

The red lines in Figure \ref{fig:pm} represent the effects of irradiation on the transient UCXBs.
Obviously, the radiation results in
a great enhance of mass-transfer rate. The duration of the enhance for the transient UCXBs
with 0.02, 0.018 and 0.015 $M_\odot$ WDs
 is only several months and even several days, which is similar to
 the timescale of irradiated WD reaching a new thermodynamic equilibrium.
 For UCXB transients, the timescale outburst is uncertainty.
 The outburst duration of XTE J1751-305 is about 15 days\citep{ Markwardt2002},
 it is about 100 days for XTE J1807-294\citep{ Falanga2005},
 while the decrease in the bolometric X-ray flux of 4U 1626-67 has lasted for about 30 yr\citep{Krauss2007}.
The outburst of UCXB transients originates from the thermal disk instability,
and theoretically it can last several months\citep[e. g.,][]{ Lasota2001}.
During the outburst, WDs are irradiated by high X-ray luminosity.
This irradiation can trigger a great enhance of mass-transfer rate.
For UCXBs with 0.02, 0.018 and 0.015 $M_\odot$ WDs, the duration of this enhance is
comparable with the timescale of the outburst, even shorter than the latter.
Therefore, the effects of the irradiation on these UCXBs are covered
in the outburst, and are not significant on observations.   However, it can last hundreds of
years for the transient UCXBs with 0.012 and 0.01 $M_\odot$ WDs, which are longer
than the theoretical and observational durations of outburst.
Therefore, from the perspective of observers, they are persistent sources.
If we increase
$f_{\rm irr}^0$ via varying $\alpha_{irr}$ from 0.01 to 0.1, the mass-transfer rates are enhanced more highly.
The duration of the outburst will shorten because the high mass-transfer rate results in orbit period widening more rapidly.

The positions of the three persistent UCXBs (XTE J0929-314, 4U 1916-05 and SWIFT J1756.9-2508) in Figure \ref{fig:pm}
are consistent with the transient sources.
However, the observations of long-term X-ray luminosity behavior for UCXBs have lasted about 20 years\citep{Haaften2012b}.
It is not long enough to judge transient sources with a outburst of hundreds of years.
We consider that due to irradiation the outbursts of the transient UCXBs are lengthened to hundreds of years.
From observations, a transient UCXB becomes a persistent source.

\begin{table*}
\centering
 %\begin{minipage}{170mm}
  \caption{The physical parameters for irradiation model.
  The first and second columns give the masses and radii of WDs, respectively. Column three shows
  the orbital periods. The forth column is the mass-transfer rates, and the X-ray luminosities produced by
  the accretion NS are in column five. The irradiation flux on the surfaces of WDs and the irradiation
  depth are in the sixth and seventh columns, respectively. }
  \tabcolsep1.0mm
 \begin{tabular*}{170mm}{llccccc}
\cline{1-7}
 $M_{\rm WD}(M_\odot)$&$R_{\rm WD}(R_\odot)$&$P_{\rm orb}({\rm Min})$&$\log \dot{M}_{\rm RL}$($M_\odot{\rm yr^{-1}}$)
 & $L_{\rm X}$$({\rm erg s^{-1}})$&$f_{\rm irr}^0$$({\rm erg s^{-1} cm^{-2}})$&$r_{\rm irr}({\rm  cm})$\\
 \cline{1-7}    %%%&
 &&&Persistent&&&\\
 0.1&0.034&10&-7.6&$7.2\times10^{37}$&$3.0\times10^{14}$&10\\
 0.08&0.035&12&-7.9&$4.0\times10^{37}$&$1.7\times10^{14}$&20\\
 0.06&0.036&14&-8.2&$1.6\times10^{37}$&$5.3\times10^{13}$&50\\
 0.05&0.037&17&-8.5&$8.7\times10^{36}$&$2.3\times10^{13}$&80\\
 0.04&0.039&20&-8.9&$3.8\times10^{36}$&$8.0\times10^{12}$&100\\
 &&&Transient&&&\\
 0.02&0.042&32&-9.9&$1.0\times10^{37}$&$1.2\times10^{13}$&$9.0\times10^3$\\
 0.018&0.043&34&-10.0&$1.0\times10^{37}$&$1.1\times10^{13}$&$1.3\times10^4$\\
 0.015&0.044&40&-10.2&$1.0\times10^{37}$&$9.5\times10^{12}$&$2.0\times10^4$\\
 0.012&0.045&44&-10.5&$1.0\times10^{37}$&$7.6\times10^{12}$&$2.5\times10^4$\\
 0.01&0.047&55&-10.8&$1.0\times10^{37}$&$6.4\times10^{12}$&$2.7\times10^4$\\
 \cline{1-7}
 \label{tab:irr}
\end{tabular*}
%\end{minipage}
\end{table*}

\section{Conclusions}
We investigate the influences of irradiation on UCXBs' evolution.
The irradiation hardly affects the evolution of
persistent sources and the transient sources during quiet phase.
However, it can trigger high mass-transfer rates in the transient sources during outburst phase.
Especially, for these sources with WDs whose masses are less than $\sim0.012 M_\odot$,
the high mass-transfer rates can last hundreds of years.
The three UCXBs, XTE J0929-314, 4U 1916-05 and SWIFT J1756.9-2508, have orbital periods of 40-60 minutes.
Theoretically, they should be transient sources, while they are persistent sources during observations of about 20 yr.
Based on the positions of UCXBs evolution, we suggest that the three persistent UCXBs should be transient sources,
and their outburst can last hundreds of years due to the irradiation of
high X-ray luminosity ($10^{37}$ erg s$^{-1}$) on the extremely low-mass WD ($\preceq0.012 M_\odot$).

\section*{Acknowledgments}
This work was supported by the National Natural Science Foundation
of China under Nos. 11473024, 11363005, 11763007, 11503008, 11365022,
and the XinJiang Science
Fund for Distinguished Young Scholars under No. QN2016YX0049.
%\newpage %Just because of unusual number of tables stacked at end
\bibliography{lglapj}
%\begin{thebibliography}{99}
%\include{lgl.bib}
%\end{thebibliography}

\label{lastpage}

\end{document}